\documentclass[a4paper,fleqn,usenatbib]{mnras}

\usepackage{newtxtext,newtxmath}
\usepackage[T1]{fontenc}
\usepackage{ae,aecompl}

\usepackage{graphicx}	% Including figure files
\usepackage{amsmath}	% Advanced maths commands
\usepackage{amssymb}	% Extra maths symbols

\def\note #1]{{\bf #1]}}

\title[Uncertainties of the Solar Neutrino Flux]{A Semi-Analytical Computation of the Theoretical Uncertainties of the Solar Neutrino Flux}

\author[A. C. S. J\o rgensen, J. Christensen-Dalsgaard]{
Andreas C. S. J\o rgensen$^{1}$\thanks{E-mail: acsj@mpa-garching.mpg.de}
and J{\o}rgen Christensen-Dalsgaard$^{2}$
\\
$^{1}$Max Planck Institut f\"ur Astrophysik, Karl-Schwarzschild-Strasse 1, 85748 Garching, Germany\\
$^{2}$Stellar Astrophysics Centre, Department of Physics and Astronomy, Aarhus University, Ny Munkegade 120, DK-8000 Aarhus C, Denmark\\
}

\date{Accepted XXX. Received YYY; in original form ZZZ}

\pubyear{2016}

\begin{document}
\label{firstpage}
\pagerange{\pageref{firstpage}--\pageref{lastpage}}
\maketitle

% Abstract of the paper
\begin{abstract}
We present a comparison between Monte Carlo simulations and a semi-analytical approach that reproduces the theoretical probability distribution functions of the solar neutrino fluxes, stemming from the $pp$, $pep$, $hep$, $^7\mathrm{Be}$, $^8\mathrm{B}$, $^{13}\mathrm{N}$, $^{15}\mathrm{O}$, and $^{17}\mathrm{F}$ source reactions. We obtain good agreement between the two approaches. Thus, the semi-analytical method yields confidence intervals that closely match those found, based on Monte Carlo simulations, and points towards the same general symmetries of the investigated probability distribution functions. Furthermore, the negligible computational cost of this method is a clear advantage over Monte Carlo simulations, making it trivial to take new observational constraints on the input parameters into account.
\end{abstract}

% Select between one and six entries from the list of approved keywords.
% Don't make up new ones.

%%%%%%%%%%%%%%%%%%%%%%%%%%%%%%%%%%%%%%%%%%%%%%%%%%

%%%%%%%%%%%%%%%%% BODY OF PAPER %%%%%%%%%%%%%%%%%%

\section{Introduction}

Over the last century, the progress in particle and nuclear physics has contributed to a thorough understanding of the interior structure of the Sun, and well-constrained solar models, on the other hand, have been used to shed light on the employed input physics. Thus, solar models and observations have led to a better understanding of neutrinos and vice versa \citep[cf.][]{jcd1996, Bahcall1998}.

To compare model predictions with observations, it is essential to establish theoretical parameter estimates as well as thorough uncertainties for the investigated parameters. One way to obtain these uncertainties, in the case of the theoretical solar neutrino fluxes from different source reactions, is to map the associated probability distribution functions based on a Monte Carlo simulation. Such an analysis has been performed by \cite{Bahcall2006}, including 10,000 standard solar models and 21 relevant input parameters. Recently, \cite{Vinyoles2017} have published yet another Monte Carlo analysis based on yet another 10,000 standard solar models, including updated solar input physics.

While a Monte Carlos analysis is reliable, it is also cumbersome and requires large amounts of computing time. In the present paper, we present a semi-analytical method that is computationally light and capable of reproducing the correct theoretical probability distribution functions.

Our work will be presented in the following order: in section~\ref{sec:method}, we elaborate on the method used, and specify both the considered neutrino source reactions and the input physics employed. In section~\ref{sec:results}, we present the results of our analysis and compare these to the results of the Monte Carlo analyses by \cite{Bahcall2006} and \cite{Vinyoles2017}.

\section{Method} \label{sec:method}

As pointed out by \cite{Bahcall2005}, \cite{Pena2008} and \cite{Serenelli2013}, the logarithm of the relative change in the predicted neutrino flux of any source reaction depends approximately linearly on the logarithm of the relative change in the input parameters. In other words, the predicted neutrino flux shows a power-law dependence on the input parameters:
\begin{equation}
\frac{\phi_i}{\phi_i(0)} = \prod_{j=1}^{N}\left( \frac{\beta_j}{\beta_j(0)}\right)^{\alpha(i,j)}, \quad \alpha(i,j)= \frac{\partial\ln \phi_i}{\partial\ln \beta_j}, \label{eq:polyparams}
\end{equation}
Here $\phi_i$ denotes the neutrino flux of the $i$th source reaction, $\beta_j$ refers to the $j$th input parameter and $\phi_i(0)$ and $\beta_j(0)$ are the corresponding values of the flux and the input parameter, respectively, for the chosen reference model.

We have tested the applicability of the approximation given by equation~(\ref{eq:polyparams}), using the Aarhus Stellar evolution code, ASTEC \citep[cf.][]{jcd2008}. We find the approximation to hold true, even when the input parameters are changed by several standard deviations, and obtain results that are in good agreement with the literature \citep[cf.][]{Joergensen2015}. 

Having established an analytical expression for the neutrino flux of each source reaction, and knowing the uncertainties of the input parameters, it is possible to reproduce the probability distribution function (PDF) of each of the considered neutrino fluxes, as we will show in the following. The idea is to select values of $\beta_j$, based on the corresponding probability distributions of the input parameters, and to calculate $\phi_i$ for each selected set of $\beta_j$, using equation~(\ref{eq:polyparams}). If the number of sets of $\beta_j$ drawn from the multivariate distribution function of the input parameters is sufficiently large, the number density of sets drawn from a given region of the corresponding parameter space will reflect the probability density of the said region. Just as in a Monte Carlo analysis, the resulting distribution of $\phi_i$ will consequently reflect the PDF of the neutrino flux. However, as opposed to a Monte Carlo analysis, no further solar models have to be computed to obtain the fluxes. A handful of models suffices to establish $\alpha(i,j)$.

Assuming that all input parameters are uncorrelated and normally distributed, the obtained $68.27\,\%$ confidence intervals for the neutrino fluxes will be well-approximated by the law of propagation of error:
\begin{equation}
\sigma\left(\frac{\phi_i}{\phi_i(0)} \right) \approx \sqrt{\sum_j \left(\alpha(i,j)\sigma(\beta_j)\right)^2}. \label{eq:lawerr}
\end{equation}
This procedure has been used as the standard approach by other authors as suggested by \cite{Villante2014}. However, since the relevant composition variables follow log-normal distributions, as discussed in Subsection~\ref{subsec:Input}, the assumptions underlying eq.~(\ref{eq:lawerr}) are not fulfilled. 

Other authors have suggested to estimate the uncertainties of the neutrino fluxes based on fractional uncertainties in the input parameters \citep[cf.][]{Bahcall2005}:
\begin{equation}
\frac{\Delta \phi_{i,j}}{\phi_i} = \left(1+\frac{\Delta \beta_j}{\beta_j} \right)^{\alpha(i,j)}-1. \label{eq:otherauth}
\end{equation}
However, as opposed to the method elaborated in the present paper, no clear statistical interpretation of the uncertainty given by equation~(\ref{eq:otherauth}) exists.

\subsection{Output: Neutrino Fluxes}

In accordance with \cite{Bahcall2006} and \cite{Vinyoles2017}, we distinguish between eight different neutrino fluxes: the fluxes from five neutrino source reactions in the PP-chain and three neutrino source reactions that are involved in the CNO cycle. As regards the PP-chain, we look at neutrinos stemming from the $p(p,\mathrm{e^+}\nu_{\mathrm{e}})\mathrm{^2 H}$ reaction, the $p(e^- p, \nu_{\mathrm{e}})\mathrm{^2H}$ reaction, the $\mathrm{^3He}(p, \mathrm{e^+} \nu_{\mathrm{e}})\mathrm{^4He}$ reaction, the $\mathrm{^7Be}(\mathrm{e^-}, \nu_{\mathrm{e}})\mathrm{^7Li}$ reaction, and the $\mathrm{^8B}(,\mathrm{e^+}\nu_{\mathrm{e}})2\mathrm{^4 He}$ reaction. Neutrinos stemming from these reactions will be referred to as $pp$, $pep$, $hep$, $\mathrm{^7Be}$, and $\mathrm{^8B}$ neutrinos, respectively. Regarding the CNO-cycle, we include the $\mathrm{^{13}N}(,\mathrm{e^+}\nu_{\mathrm{e}})\mathrm{^{13} C}$ reaction, the $\mathrm{^{15}O}(,\mathrm{e^+} \nu_{\mathrm{e}})\mathrm{^{15} N}$ reaction, and the $\mathrm{^{17}F}(,\mathrm{e^+}\nu_{\mathrm{e}})\mathrm{^{17} O}$ reaction. Neutrinos stemming from these reactions will be referred to as $\mathrm{^{13}N}$, $\mathrm{^{15}O}$, and $\mathrm{^{17}F}$ neutrinos, respectively.

\subsection{Input Parameters} \label{subsec:Input}

In order to fascilitate an easy comparison with \cite{Vinyoles2017}, we employ 22 input parameters to characterize the models: age, surface luminosity, the element diffusion rate, two parameters ($a$ and $b$) parametrizing the opacity\citep[cf.][]{Vinyoles2017}, $S$-factors for 8 nuclear reactions ($S_{11}$, $S_{33}$, $S_{33}$, $S_{34}$, $S_{e7}$, $S_{17}$, $S_{\mathrm{hep}}$, $S_{1,14}$, $S_{1,16}$), and the abundances of 9 elements (C, N, O, Ne, Mg, Si, S, Ar, Fe).

Using ASTEC, we have also investigated the influence of four additional reactions\footnote{The $\mathrm{^{12}C}(p,\gamma)\mathrm{^{13}N}$ reaction, the $\mathrm{^{13}C}(p,\gamma)\mathrm{^{14}N}$ reaction, the $\mathrm{^{15}N}(p,\alpha)\mathrm{^{12}C}$ reaction, and the $\mathrm{^{15}N}(p,\gamma)^{16}\mathrm{O}$ reaction.} in the CNO cycle, but we found the respective $\alpha(i,j)$ to be negligibly small, which was to be expected, as $\mathrm{^{14}N}(p,\gamma)^{15}\mathrm{O}$ is the bottleneck-reaction and in accordance with \cite{Bahcall2006} \citep[cf.][]{Joergensen2015}.

For the sake of an easy comparison, we use the values for $\alpha(i,j)$ that were extracted from the models that enter the Monte Carlo analysis published\footnote{Tables containing the derived values of $\alpha(i,j)$ can be found on \url{http://www.ice.csic.es/personal/aldos/Solar_Data.html}.} by \cite{Vinyoles2017}. We also use the corresponding uncertainties listed in the bulk text and Table~1--3 of the same article.

The large discrepancies between the solar composition suggested by different authors has led to debate regarding the associated uncertainties. \cite{Bahcall2006} therefore distinguish between the so-called optimistic uncertainties recommend by \cite{Asplund2005} and the so-called conservative (historical) uncertainties. These conservative uncertainties are based on the difference in the abundances presented by \cite{Grevesse1998} and \cite{Asplund2005} and hence include both statistical and systematic errors. The discrepancies between the determination of the solar surface composition, leading to the conservative uncertainties employed by \cite{Bahcall2006}, are clearly a serious concern for determinations of the model neutrino fluxes and their PDFs. However, we note that since this discrepancy is not of a statistical nature it cannot, strictly speaking, be included in the assumed probability distribution of the input parameters. A more rigorous approach is probably to regard models based on the different composition determinations as distinct groups of models, each of which can be compared with the observations. However, a detailed discussion of this matter is beyond the scope of the present article, as our purpose it merely to compare our method with Monte Carlo simulations. Based on the arguments above, we simply employ the spectroscopic uncertainties recommended by \cite{Asplund2009} and \cite{Grevesse1998}, respectively.

To compute the eight investigated neutrino fluxes, it is necessary to know the probability distribution functions of each of the varied input parameters. For simplicity, we assume all input parameters to be normally distributed and thus interpret the uncertainties listed in \cite{Vinyoles2017} as standard deviations. It is worth to stress that the validity of the method presented in this paper does not depend on the assumption that the input parameters are normally distributed. Other probability distributions from which to draw the samples of input parameters may be chosen.

Furthermore, we assume all estimates of $\beta_j(0)$ and the corresponding standard deviations to be uncorrelated, as they follow from different observations and experiments. Note that we do not state that the model parameters are uncorrelated. Obviously, they are correlated. However, the observational constraints are not.

As regards the solar composition, the logarithms of the abundances that are generally quoted in the literature are assumed to follow Gaussian distributions. Consequently, the relative changes in the individual abundances, to which the corresponding values of $\alpha(i,j)$ refer, follow log-normal distributions. We therefore draw the logarithms of the abundances from normal distributions as described above and convert these into relative changes in the abundances before computing the associated changes in the neutrino fluxes.

\section{Results} \label{sec:results}

\begin{figure}
	% To include a figure from a file named example.*
	% Allowable file formats are eps or ps if compiling using latex
	% or pdf, png, jpg if compiling using pdflatex
	\includegraphics[width=\columnwidth]{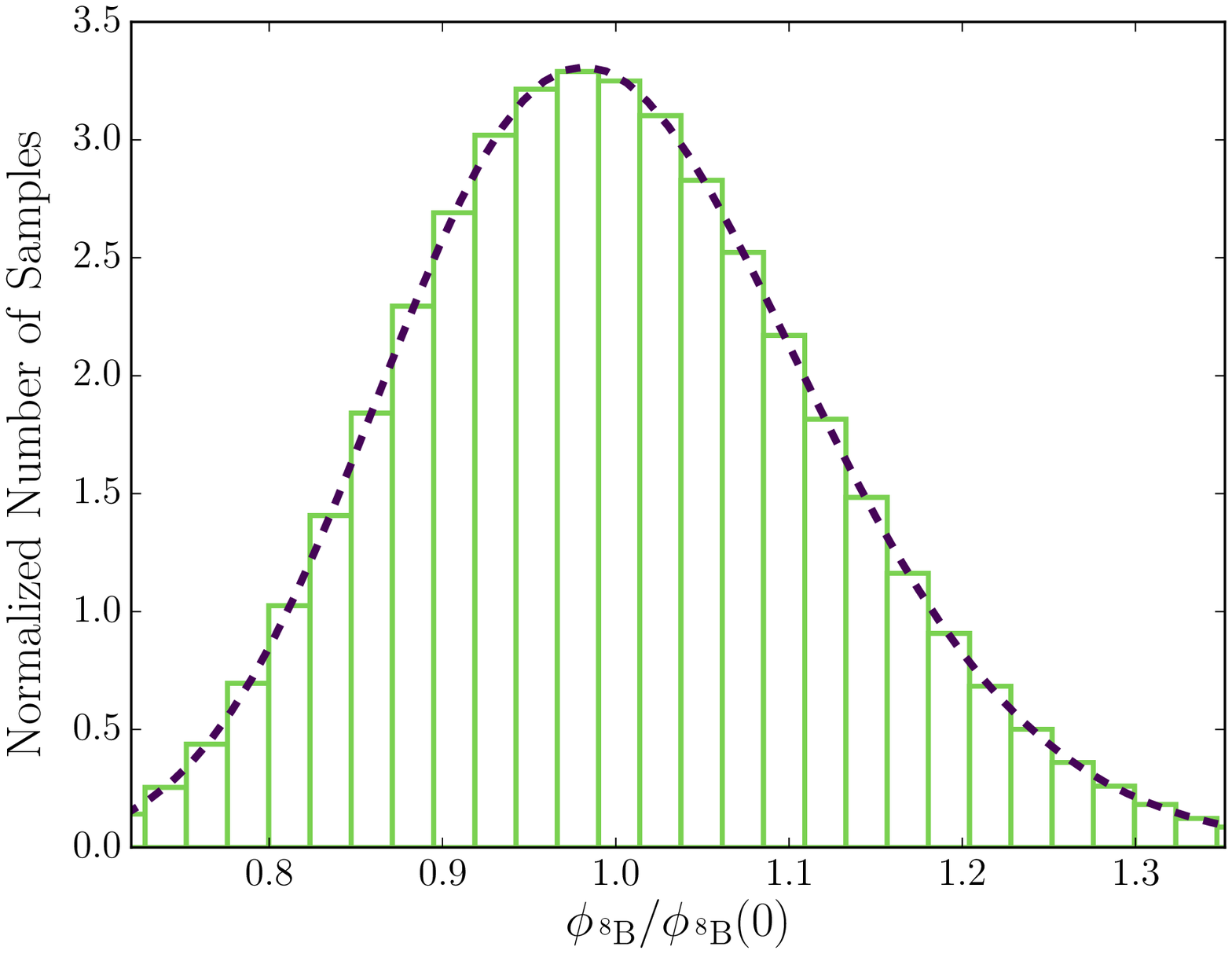}
    \caption{Histograms of the evaluated normalized probability distribution of the relative change in the $\mathrm{^8B}$ neutrino flux, based on $10^6$ combinations of input parameters drawn randomly from the corresponding normal distributions. We used AGSS09. The plot shows the distribution between the 0.5th and 99.5th percentile, and includes the best-fitting log-normal distribution.}
    \label{fig:B8Dist}
\end{figure}

Fig.~\ref{fig:B8Dist} shows the probability distributions obtained for the $\mathrm{^8B}$ neutrino flux, based on $10^6$ combinations of input parameter drawn randomly from the corresponding multivariate Gaussian distribution. Thus, the presented PDF corresponds to the output of a Monte Carlo analysis involving one million standard solar models.

Table~\ref{tab:resultsTable} summarizes the computed probability distribution functions of the eight investigated neutrino fluxes. 

\begin{table*}
	\centering
	\caption{Uncertainties of the eight investigated neutrino fluxes, using the abundances by GS98 and AGSS09. Columns labelled "Percentiles" list the associated 15.87th, 50th and 84.13th percentiles of $\phi_i/\phi_i(0)$, summarizing the empirical probability distribution. $\sigma_-$ and $\sigma_+$ denote the relative difference between the 15.87th and the 50th percentiles and between the 84.13th and 50th percentiles, respectively. Columns labelled "Distribution" list, whether the associated probability distribution is well-approximated by a normal distribution or a log-normal distribution, according to a Kolmogorov-Smirnov~test, at a 5$\,$\% significance level. $10^6$ combinations of input parameters were employed.}
	\label{tab:resultsTable}
	\begin{tabular}{l | c c c c c c c c} % four columns, alignment for each
		\hline
&  \multicolumn{3}{c}{\underline{GS98}} & \multicolumn{3}{c}{\underline{AGSS09}} \\[2 pt]
		Flux & Percentiles & $[-\sigma_-;\sigma_+]$ ($\%$) & Distribution & Percentiles & $[-\sigma_-;\sigma_+]$ ($\%$) & Distribution \\[2 pt]
		\hline
 $pp$ & $[0.994;1.000;1.006]$ & $[-0.60;0.62]$ & Neither & $[0.994;1.000;1.006]$ & $[-0.58;0.59]$ & Neither \\[2 pt]
 $pep$ & $[0.990;1.000;1.011]$& $[-1.00;1.02]$ & Neither & $[0.991;1.000;1.010]$ & $[-0.95;0.97]$ & Neither \\[2 pt]
 $hep$ & $[0.697;1.002;1.306]$ & $[-30.4;30.4]$ & Neither & $[0.699;1.001;1.305]$ & $[-30.2;30.4]$ & Neither \\[2 pt]
 $\mathrm{^7Be}$ & $[0.929;0.997;1.070]$& $[-6.87;7.23]$ & Log-normal & $[0.928;0.997;1.070]$ & $[-6.92;7.31]$ & Log-normal \\[2 pt] 
 $\mathrm{^8B}$ & $[0.880;0.994;1.122]$ & $[-11.5;12.8]$ & Log-normal & $[0.879;0.994;1.122]$ & $[-11.6;12.9]$ & Log-normal \\[2 pt]
 $\mathrm{^{13}N}$ & $[0.852;0.993;1.157]$ & $[-14.3;16.4]$ & Log-normal & $[0.867;0.994;1.138]$ & $[-12.8;14.5]$ & Log-normal \\[2 pt]
 $\mathrm{^{15}O}$ & $[0.837;0.992;1.173]$ & $[-15.7;18.2]$ & Log-normal & $[0.848;0.993;1.158]$ & $[-14.5;16.6]$ & Log-normal \\[2 pt]
 $\mathrm{^{17}F}$ & $[0.808;0.991;1.213]$ & $[-18.5;22.4]$ & Log-normal & $[0.825;0.992;1.189]$ & $[-16.9;19.9]$ & Log-normal \\[2 pt]
		\hline
	\end{tabular}
\end{table*}

\subsection{Comparison with Monte Carlo Analyses} \label{subsec:monte}

As mentioned in the introduction, \cite{Bahcall2006} and \cite{Vinyoles2017} presented probability distributions for solar neutrino fluxes, based on Monte Carlo analyses, including 10,000 standard solar models. Broadly speaking, both Monte Carlo analyses led to results that are consistent with the confidence intervals summarized in Table~\ref{tab:resultsTable}; a detailed comparison turns out to be fruitful.

Firstly, as can be seen from Table~\ref{tab:resultsTable}, all confidence intervals are asymmetric. Such asymmetries are also found in the cited Monte Carlo analyses. According to \cite{Bahcall2006}, the asymmetric probability distributions of the $^8\mathrm{B}$ and the CNO neutrino fluxes are more well-approximated by log-normal distributions, while the remaining fluxes follow Gaussian distributions. The asymmetries of the former are more pronounced than in the present article. \cite{Vinyoles2017}, on the other hand, finds the asymmetries to be rather small in all cases, and ascribes Gaussian distributions to all eight neutrino fluxes.

These observed asymmetries of the probability distributions of the neutrino fluxes are largely due to the fact that the composition variables follow log-normal distributions. Thus, in accordance with \cite{Bahcall2006}, we find the asymmetries to be most pronounced for the neutrino source reactions that are involved in the CNO-cycle.

\begin{figure}
	% To include a figure from a file named example.*
	% Allowable file formats are eps or ps if compiling using latex
	% or pdf, png, jpg if compiling using pdflatex
	\includegraphics[width=\columnwidth]{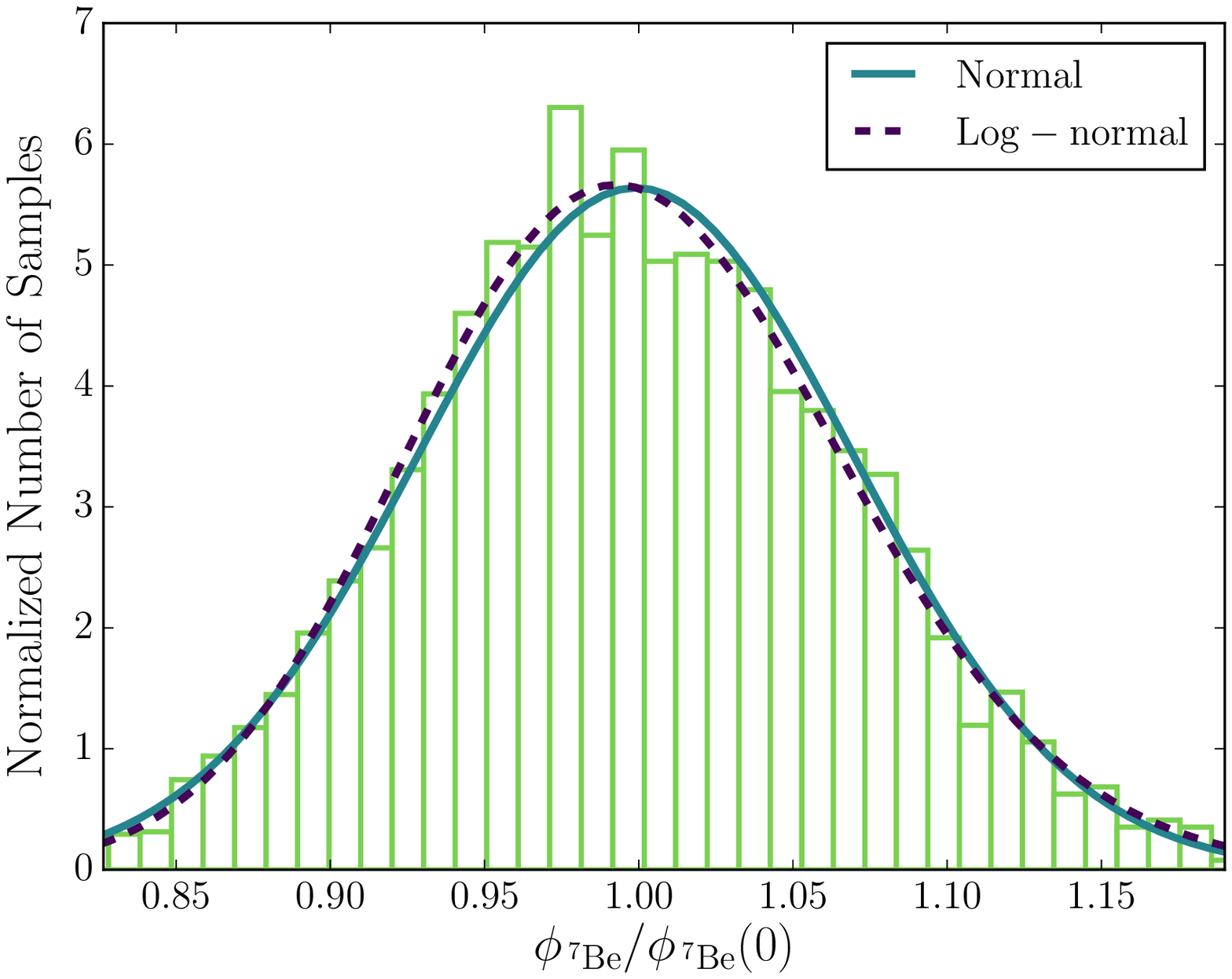}
    \caption{Histograms of the evaluated normalized probability distribution of the relative change in the $\mathrm{^7Be}$ neutrino flux, based on 5000 combinations of input parameters drawn randomly from the corresponding normal distributions. We used AGSS09. The plot shows the distribution between the 0.5th and 99.5th percentile, and includes the best-fitting normal and log-normal distributions.}
    \label{fig:B7Dist}
\end{figure}

In order to access the null hypotheses that the empirical probability distributions obtained, using our method, likewise correspond to either a normal or a log-normal distribution, at a 5$\,$\% significance level, we have employed a Kolmogorov-Smirnov test. In all cases, we have been able to reject the hypothesis that the empirical distribution is well approximated by a normal distribution, while some of the fluxes still seem to follow log-normal distributions. The results are specified in Table~\ref{tab:resultsTable}. This being said, while our analysis is based on $10^6$ samples from the relevant parameter space, \cite{Bahcall2006} and \cite{Vinyoles2017} only include 5000 models for two different solar compositions. We have thus rerun our analysis with 5000 samples and could in several cases no longer reject either of the stated null hypotheses (cf. Fig.~\ref{fig:B7Dist}). Thus, our analysis implies that a Monte Carlo analysis, based on a few thousand solar models, may lead to the conclusion that the probability distributions are well-approximated by normal and log-normal distributions, while these hypotheses can be discarded in many cases, when including more models.

Secondly, although we find the asymmetries to be more pronounced than \cite{Vinyoles2017}, the absolute values of the obtained 15.87th and 84.13th percentiles in Table~\ref{tab:resultsTable} match the results listed in Table~6 of \cite{Vinyoles2017} quite well. This holds true for both compositions considered in this paper: \cite{Grevesse1998} and \cite{Asplund2009}. In Table~\ref{tab:resultsTable}, these compositions are abbreviated as GS98 and AGSS09, respectively.

In comparison, the confidence intervals listed in \cite{Bahcall2006} are slightly broader, which is largely due to the updated reaction rates, i.e. the fact that the uncertainties on the relevant $S$-factors have been reduced significantly since 2006 \citep{Adelberger1998,Adelberger2011}. Thus, when using values of $\alpha(i,j)$ and $\sigma(\beta_j)$ that correspond to the assumptions made in \cite{Bahcall2006}, we arrive at confidence intervals that lie close to the results obtained by these authors.  

Any changes in $\beta_j(0)$ and $\sigma(\beta_j)$ necessitate a recalculation of the neutrino probability distributions. While a recomputation of a Monte Carlo analysis is rather cumbersome, updating the probability distributions of the neutrino fluxes, to take new observational constraints into account, can be achieved at a low computational cost, using the method presented in this paper. Severe changes in the input physics that led to changes in $\beta_j(0)$ may affect $\alpha(i,j)$. Hence, the values listed in the literature differ, depending on the underlying assumptions, such as the composition. However, only an handful of models are needed to reestablish $\alpha(i,j)$.

\section{Conclusion}

We have presented a computationally-light semi-analytical approach to evaluate the theoretical probability distributions of the solar neutrino flux for different source reactions. This approach is based on the linear response of the logarithm of the predicted flux to changes in the logarithm of different input parameters. As pointed out by \cite{Haxton2008} and \cite{Pena2008}, this linear relationship can be expressed in a single parameter, $\alpha(i,j)$.

The results obtained from this semi-analytical approach are in good agreement with results obtained from Monte Carlo analyses and reveal the same general symmetries of the PDFs of the neutrino fluxes. Hence, this method reliably provide confidence intervals at any confidence level. Furthermore, the low computational cost of the presented method is a clear advantage over a Monte Carlo analysis. Thus, $\alpha(i,j)$ can be evaluated based on only a handful of solar models. Moreover, as the computational costs of computing the probability distributions is negligible, keeping the uncertainties up to date and including new observational constraints on input parameters is trivial. 

\section*{Acknowledgements}
We record our gratitude to R.~Handberg and A.~Serenelli for useful discussions and insights. 
Funding for the Stellar Astrophysics Centre is provided by 
the Danish National Research Foundation (Grant DNRF106).
The research was supported by the ASTERISK project
(ASTERoseismic Investigations with SONG and Kepler)
funded by the European Research Council (Grant agreement no.: 267864).

%%%%%%%%%%%%%%%%%%%%%%%%%%%%%%%%%%%%%%%%%%%%%%%%%%

%%%%%%%%%%%%%%%%%%%% REFERENCES %%%%%%%%%%%%%%%%%%

% The best way to enter references is to use BibTeX:

%\bibliographystyle{mnras}
%\bibliography{example} % if your bibtex file is called example.bib

% Alternatively you could enter them by hand, like this:
% This method is tedious and prone to error if you have lots of references

%%%%%%%%%%%%%%%%%%%%%%%%%%%%%%%%%%%%%%%%%%%%%%%%%%

%%%%%%%%%%%%%%%%% APPENDICES %%%%%%%%%%%%%%%%%%%%%

%\appendix

%\section{Some extra material}

%If you want to present additional material which would interrupt the flow of the main paper,
%it can be placed in an Appendix which appears after the list of references.

%%%%%%%%%%%%%%%%%%%%%%%%%%%%%%%%%%%%%%%%%%%%%%%%%%

% Don't change these lines
\bsp	% typesetting comment
\label{lastpage}
\end{document}